\documentclass[a4paper]{IEEEtran}

\usepackage{balance}

\usepackage{graphicx} 

\usepackage[english]{babel}
\usepackage[utf8]{inputenc}
\usepackage{mathtools}
\usepackage{tikz}
\usepackage{amsmath,amssymb,amsfonts}
\usepackage{pdfpages}

\usepackage{textcomp}
\usepackage{xcolor}
\def\BibTeX{{\rm B\kern-.05em{\sc i\kern-.025em b}\kern-.08em
    T\kern-.1667em\lower.7ex\hbox{E}\kern-.125emX}}

\usepackage{afterpage}
\usepackage{bm}
\usepackage{color}
\usepackage[font=sc]{caption}

\usepackage{listing}
\usepackage{listings}
\usepackage{fancyhdr}
\usepackage{fancyvrb} 
\usepackage{float}
\usepackage{framed}
\usepackage{enumerate}
\usepackage[shortlabels]{enumitem}
\usepackage{multirow}
\usepackage{hhline}
\usepackage{array}
\usepackage{epstopdf}
\usepackage{pgfplots}
\usepackage{subfigure}

\usepackage{dblfloatfix}    

\usetikzlibrary{matrix,chains,positioning, decorations.pathreplacing,arrows,plotmarks}
\usetikzlibrary{shapes}
\usetikzlibrary{calc} 
\pgfplotsset{compat=newest}

\usetikzlibrary{shapes.geometric, arrows}
\tikzstyle{db} = [rectangle, minimum width=0.4\columnwidth, minimum height=0.2\columnwidth, text centered, draw=black, fill=blue!20]
\tikzstyle{file} = [rectangle, rounded corners, minimum width=0.3\columnwidth, minimum height=0.1\columnwidth, text centered,
draw=black, fill=green!20]
\tikzstyle{dir} = [circle, minimum size=0.2\columnwidth, fill=red!20, draw=black]
\tikzstyle{arrow} = [thick, ->, >=stealth]
\tikzstyle{user} = [rectangle, minimum width=0.2\columnwidth, minimum height=0.05\columnwidth, draw=black, fill=black!20]

\usepackage[ruled, vlined]{algorithm2e}

\IEEEoverridecommandlockouts 
\usepackage[OT4,T1]{fontenc}
\usepackage{url}
\usepackage{epstopdf}

\usepackage{cite} 

\newcommand\MyBox[2]{
  \fbox{\lower0.75cm
    \vbox to 1.7cm{\vfil
      \hbox to 1.7cm{\hfil\parbox{1.4cm}{#1\\#2}\hfil}
      \vfil}%
  }%
}

\title{Application of Audio Fingerprinting Techniques for Real-Time Scalable Speech Retrieval and Speech Clusterization}
\author{
\IEEEauthorblockN{Kemal Altwlkany $^{1,2}$, Sead Delalić $^{1,2}$, Adis Alihodžić $^{1}$, Elmedin Selmanović $^{1}$, Damir Hasić $^{1}$
}

\IEEEauthorblockA{
$^1$Faculty of Science, University of Sarajevo, Bosnia and Herzegovina\\
$^2$Infobip, Sarajevo, Bosnia and Herzegovina\\
  \{kemal.altwlkany, delalic.sead, adis.alihodzic, eselmanovic, d.hasic\}@pmf.unsa.ba
}
}

\begin{document}
\maketitle

\begin{abstract}
Audio fingerprinting techniques have seen great advances in recent years, enabling accurate and fast audio retrieval even in conditions when the queried audio sample has been highly deteriorated or recorded in noisy conditions. Expectedly, most of the existing work is centered around music, with popular music identification services such as Apple's Shazam or Google's Now Playing designed for individual audio recognition on mobile devices. However, the spectral content of speech differs from that of music, necessitating modifications to current audio fingerprinting approaches. This paper offers fresh insights into adapting existing techniques to address the specialized challenge of speech retrieval in telecommunications and cloud communications platforms. The focus is on achieving rapid and accurate audio retrieval in batch processing instead of facilitating single requests, typically on a centralized server. Moreover, the paper demonstrates how this approach can be utilized to support audio clustering based on speech transcripts without undergoing actual speech-to-text conversion. This optimization enables significantly faster processing without the need for GPU computing, a requirement for real-time operation that is typically associated with state-of-the-art speech-to-text tools.
\end{abstract}

\providecommand{\keywords}[1]
{
  \textbf{\textit{Keywords---}} #1
}


\section{Introduction}

\IEEEoverridecommandlockouts\IEEEPARstart{A}{udio} fingerprinting is a technique that generates unique \textit{fingerprints} from audio content, enabling their distinct identification from one another. Although the exact definition of a fingerprint varies depending on the concrete method of fingerprinting used, the key concepts remain the same. The problem of fingerprinting is especially interesting as there is a certain duality regarding it: the fingerprints should be robust enough in order to recognize distorted and degraded versions of the same audio content, but sensitive enough to distinguish between different audio contents themselves \cite{seo2005audio}.

The broader public is probably more familiar with popular music identification services such as: Shazam (Apple), Now Playing (Google), Musixmatch, ACRCloud, Midomi, SoundHound and many other. Some of these services have made their underlying audio fingerprinting techniques publicly available \cite{wang2003industrial,baluja2008waveprint,gfeller2017now}. The modus operandi of music identification services is more or less similar: the end-user records a very small part of a song which is playing in their environment and the system will provide them with the name of the song, the song's artist and other useful metadata.

Existing work on audio fingerprinting has mostly devoted its attention to \textit{music} instead of \textit{speech}, which is quite reasonable. First of all, popular music identification services make more sense, as it is more likely to encounter an unknown song playing in a shopping mall, cafe or bus station, rather than a podcast or radio show. Second of all, when considering speech retrieval, it is often desirable to retrieve a collection of relevant audio files based on the spoken words and meaning, rather than the actual \textit{audio content}, as in the waveform of the utterances \cite{glavitsch1995first,lee2015spoken}. However, such systems for speech retrieval require specific attention, as it is not an easy task to obtain text from speech, with state-of-the-art quality.

\subsection{Motivation}
Infobip Ltd. is an international company that specializes in omnichannel communication, enabling clients to interact with their customers using email, SMS, voice, video and similar platforms \cite{infobip}. Of particular interest for this paper is voice communication facilitated through Infobip, which includes both Voice over IP (VoIP), as well as public switched telephone network (PSTN) calls. 

While performing work at Infobip Ltd. \cite{infobip}, some of the authors of this paper have encountered a problem to which speech retrieval seemed as the most appropriate solution. What follows is a brief description of the problem at hand, as it was the main motivation of this work.

\subsubsection{A Glance at SIP and Early Media}
The Session Initiation Protocol (SIP) is an application-layer protocol that enables creating, modifying, and terminating sessions with one or more participants \cite{rfc3261}. These sessions include Internet telephone calls, multimedia distribution, and multimedia conferences.

RFC 3960 defines Early media \cite{rfc3960} within SIP, as media (e.g., audio and video) that is exchanged before a particular session is accepted by the called user. A bold simplification of these terms can be formulated as: before establishing a call, the caller and callee can exchange media, referred to as early media. Typical examples of early media are ringing tone and announcements.

The announcements exchanged as early media are of particular interest, as they often contain additional information about the call status, e.g. they can contain a voicemail, or indicate whether the dialed user is busy, unavailable, out of network coverage, that a bad gateway was selected, etc. This information should be encapsulated within SIP using SIP status codes \cite{rfc3261}, but it can happen that the information is omitted or that an incorrect SIP response code has been returned. In these cases, analyzing the content of early media can be useful:
\begin{itemize}
    \item The results of analysis can be used immediately/online to perform call actions such as redialling the callee, or transfering the call to another gateway.
    \item The results of analysis can be used a posteriori/offline in order to gain better insights into the network performance and potentially discover issues with the network.
\end{itemize}

\subsubsection{Problem Definition}
Early media files described in the previous subsection vary significantly, depending on the country and specific data center (DC) in which the calls are being established. Early media is grouped into clusters, per DC, with each cluster resembling a state of importance: e.g. one cluster can resemble all early media for which the dialed number was busy, another cluster can resemble all early media which contains a voicemail, etc. These clusters are defined by business use-cases and their meaning does not affect the main research premise.

In this paper, a solution will be proposed that classifies received early media without an accompanying SIP response code. This classification will be performed by identifying each early media file using audio fingerprinting techniques and then assigning the file to the correct cluster. The procedure must be done in real-time, preferably as soon as possible.

\section{Related Work}

Audio fingerprinting is the primary technique employed in this paper. However, as will be demonstrated shortly, speech-to-text (STT) or automatic speech recognition (ASR) at state-of-the-art quality is a necessity as well. Thus, relevant work from both spectrums is reviewed.

\subsection{Audio Fingerprinting}

Perhaps not the first paper on the given topic, but definitely a revolutionary one was the paper of Wang in which Shazam was presented \cite{wang2003industrial}. The basic idea is creating a spectrogram of the audio and filtering out most of the spectral content except for spectral peaks. The main premise is that local maxima of the spectrogram are less prone to changing their positions within the spectrogram once noise and distortions occur. Both time and frequency coordinates of peaks are used and the peaks are paired and hashed. These hashes of pairs of frequency peaks are what forms the fingerprint of an audio sample. Retrieval is performed by: recording audio in real-time, generating a live spectrogram, creating the fingerprints from the spectrogram and finally, searching the database for them.

Baluja and Covell \cite{baluja2006content,covell2007known,baluja2008waveprint} have published several papers which share the main idea of Shazam, in their approach named Waveprint. They have observed that the problem of audio retrieval if based upon a spectrogram can be thought of as a computer vision problem. After all, a spectrogram is an image, albeit a special one. Their solution is based upon the work of Ke et al. \cite{ke2005computer} in which a machine learning technique, \textit{AdaBoost} is applied. Inspired by the work of Jacobs et al. \cite{jacobs1995fast}, Baluja and Covell opted for a wavelet-based approach instead of a machine learning approach. This is a major change compared to Wang's approach \cite{wang2003industrial}, as wavelets enable multi-resolution analysis, whereas a standard spectrogram provides a fixed frequency-time resolution \cite{mallat1989theory,kovacevic2013fourier}.

The second change introduced by Baluja and Covell is the usage of logarithmically spaced frequency bins, instead of linearly spaced frequency bins. They have referenced the work of Haitsma and Kalker \cite{haitsma2002highly} as one of the most widely used audio fingerprinting systems. The approach of Haitsma and Kalker uses a logarithmical scale of frequency as well, but better accustomed to the human auditory system - the Bark scale (they also consider using the Mel scale) \cite{moore2012introduction,ballou2013handbook}.

Finally, Waveprint \cite{baluja2006content,covell2007known,baluja2008waveprint} relies on locally sensitive hashing for the retrieval process. Once a fingerprint has been generated, a locally sensitive hashing technique, Min-Hash in particular for Waveprint, is used \cite{gionis1999similarity,cohen2001finding}.

\subsection{Automatic Speech Recognition}
In their 2019 paper, Ravanelli et al. \cite{ravanelli2019pytorch} provide a nice introduction describing the status of ASR. They contribute the success of state-of-the-art ASR to several factors: advances in deep learning, speech recognition challenges such as CHiME \cite{barker2017third}, publicly available datasets such as Librispeech \cite{panayotov2015librispeech} and the research advances of open-source ASR software such as Julius \cite{lee2009recent} and Kaldi \cite{Povey_ASRU2011}, both of which are non-deep learning based.

Kaldi \cite{Povey_ASRU2011} is a framework for speech processing that relies on finite-state transducers, but offers a few pretrained models, most of which are for English and Mandarin, whereas Julius \cite{lee2009recent} provides models for Japanese.


Whisper \cite{radford2023robust} released by OpenAI uses a transformer architecture \cite{vaswani2017attention}. The authors describe the model as a general-purpose speech recognition model that can perform multilingual speech recognition, speech translation, and language identification. Whisper was trained on an immense dataset of over 680.000 hours of audio. Several variants of the model exist, which provide a trade-off between accuracy and speed: tiny, small, base, medium and large. All models are multilingual, but some do come in english only version. When benchmarked on the FLEURS dataset \cite{conneau2023fleurs}, the multilingual large model performs best for Spanish, Italian and Korean, achieving an a WER of 2.8, 3.0 and 3.1, respectively.

Meta has contributed a lot to the area of ASR, with a focus on contributing to the preserving of language diversity \cite{pratap2020massively,tjandra2023massively,pratap2023scaling}. Their latest Massively Multilingual Model (MMS) supports over a 1100 different languages for ASR and is able to perform language identification of over 4000 different languages. Similar to Whisper, it relies on the powerful Transformer architecture \cite{vaswani2017attention}, but the authors claim that it more than halves the WER of Whisper on 54 languages of the FLEURS dataset \cite{conneau2023fleurs}.

\begin{figure*}[!b]
    \centering
    \includegraphics[width=1.0\linewidth]{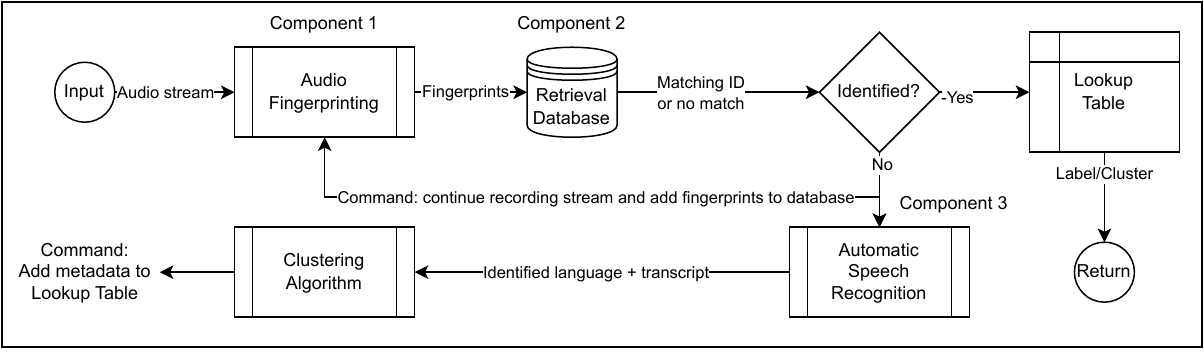}
    \caption{Diagram representing the modules and pipeline of the entire system.}
    \label{fig::system diagram}
\end{figure*}

\section{Case Study}

An overview of the entire system is provided on Fig. \ref{fig::system diagram} with some minor details/connections between components being omitted from the image for a clearer graphical representation. All components of the system are represented as logical units, not necessarily actual independent microservices and the main components have been labeled with numbers: 1, 2 and 3 on Fig. \ref{fig::system diagram}. What follows is an overview of the entire procedure and the individual components of the system. Training the system is also reviewed, as the system cannot be released into production without undergoing an initial training phase.

\subsection{Procedure}

Firstly, the audio is streamed into the system and the fingerprinting component (component 1) will start creating fingerprints of the stream in real-time. These fingerprints are forwarded to the database retrieval component (component 2) which identifies the audio by returning its unique ID or indicates that the audio does not exist in the database.

If the audio has been identified, its unique ID is forwarded to the look-up table component.  The look-up table returns the corresponding label/cluster of the audio. This is the end result, and the incoming stream of early media can be stopped as the label will contain information that has been missing due to the SIP response code not being present, e.g. the label carries information that this early media represents a voicemail, or an announcement stating that the user is busy, or unavailable, etc.

If the audio is not found, a procedure is performed in order to add it to the database, so that it can be identified in the future, should it ever be streamed again. This procedure ensures that the database is automatically expanded and kept up to date. When a file is not identified, instead of stopping the stream, the system continues receiving early media from the stream until the entire file is obtained.

Once the entire file has been recorded, it is assigned a new, unique ID. Now it is sent in parallel to both the ASR component (component 3) and the audio fingerprinting component.

The ASR component will perform language identification and speech-to-text on the newly obtained file. The identified language together with the transcript of the audio are encoded and passed to the clustering component. The task of the clustering component is to assign a cluster to the newly obtained file. This assignation process is based on both the language and the transcript.

Meanwhile, in parallel, the fingerprinting module will create fingerprints of the audio. These are paired with the file's unique ID and added to the database retrieval component.

Once both parallel tasks finish, the previously unknown file will be contained in the database. It was automatically assigned a label based on the identified language and spoken content, without the requirement of human interference. Since a digital fingerprint of it has been created, should this file be streamed again it will be identified and its corresponding label will be obtained from the retrieval database. This label carries information which is encapsulated in the content of the audio, but the system does not need to invoke any ASR in order to obtain that information.

\subsection{Training Phase}

The first step towards towards building the system is performing a \textit{training phase}. The training phase must be done independently per data center, as the content of exchanged early media heavily depends on the country and region in which calls are established.

For simplicity, the term \textit{engineer} will be used to refer to the person/group managing and distributing parts of the system which require human interference. Once a business decision has been made to deploy the system in a certain data center, the engineer gathers relevant samples of audio files to form an initial version of the dataset. As an example, this can be done by simply recording early media for a certain period.

Upon gathering an initial dataset, all files are assigned a unique ID and passed through the fingerprinting system. The unique IDs paired with the fingerprints are added to the retrieval database. At this point, the database itself can be used to remove duplicate audio files. This can be done by querying the database for the fingerprints of each file and marking files with a significant overlap of fingerprints as duplicates. Duplicate copies of the same audio file generally should not impact the accuracy of the system, but are redundant and can impact retrieval speed if not dealt with orderly.

Once the dataset has been filtered out for duplicates, each remaining file is passed through the ASR module. Transcripts of each file are obtained and clustered into groups. The clustering phase is the least strictly defined phase in the entire system and highly depends on the use case. For example, the transcripts can be clustered using a common clustering algorithm such as k-means \cite{macqueen1967some} or DBSCAN \cite{ester1996density}.

Afterwards, a keyword extraction algorithm can be applied separately per cluster, on each of the transcripts. The recommended next step is to have the engineer look into the keywords for each cluster and a few examples of full-text transcripts in order to determine the representation of the cluster. As an example, a cluster containing keywords: \textit{reached}, \textit{voicemail} and \textit{message} would indicate that this cluster represents voicemails. Once more, this step requires strong domain knowledge of the engineer. The clustering algorithm itself can be omitted, and the audio files can be grouped using any logic suitable for the desired use-case. What is important is that the outcome of this step is that each file has a cluster/label assigned to it and that the cluster is correlated with the spoken content of the file.

Once each file has been assigned a label, pairs of unique IDs and labels are added to the Lookup table module. By this the training phase has been concluded. The components can be deployed and start processing incoming requests as described in the start of this section.

\subsection{Audio Fingerprinting - Component 1}
At heart of the system lies the audio fingerprinting component. A careful study has been conducted of existing algorithms  which were reviewed in Section 2. However, none of the algorithms are tailored specifically for speech retrieval, which as stated in the introductory paragraphs is reasonable. Although similarities exist, speech differs from music, requiring special attention for audio fingerprinting. For example, speech signal is frequently interrupted by short silences and mostly consists of quasi-periodic signals \cite{wolfe2002speech}. In most speech, the equivalent of pitch varies continuously, whereas the pitch of individual notes in music is relatively stable \cite{wolfe2002speech}. Waveprint provided good performance and is a modern audio fingerprinting technique, but some modifications were made. The mean of the vocal fundamental frequencies of males is around 116 Hz, whereas it is 217 Hz for females \cite{fitch1970modal}, but Waveprint creates spectrograms ranging from 318 Hz to 3500 Hz \cite{baluja2006content,covell2007known,baluja2008waveprint}. This was adapted to speech, and the frequencies considered range from 100 Hz to 350 Hz. Using logarithmically spaced frequency bins was kept, but a common spectrogram with linearly spaced frequency bins was taken into consideration as well.

\subsection{Retrieval Database - Component 2}
The idea of using locally sensitive hashing from Waveprint \cite{baluja2006content,covell2007known,baluja2008waveprint} is retained in this paper as well. It is analyzed separately since some improvements were made and since the algorithm for retrieving fingerprints can be chosen independently of the fingerprinting algorithm. The decision made was to use Facebook AI Similarity Search (faiss) \cite{johnson2019billion}. Faiss scales incredibly well, it provides a lot of functionality and different indexes used for searching the database, all of which were found to perform well for the purposes of this paper. Another benefit of faiss is that it supports batched requests: multiple streams of early media can perform fingerprinting in parallel and then search the database via a single, batched query. 

\subsection{Automatic Speech Recognition - Component 3}
Careful readers may have observed that the entire system could be replaced with an \textit{"ideal"} ASR. This is true, the system could be replaced with an ASR solution that could perform real-time transcriptions of a multitude of languages and not require the usage of an expensive GPU to do so, as the cost of operating the system is a design constraint imposed at start.

However, speed and pricing are not the only issues with state-of-the-art ASR models for the problem at hand. Firstly, any model used must be able to identify the spoken language. Second, the labeling of audio must be performed as soon as possible, preferably within the first few seconds of the audio stream. This is often not enough time for models such as Whisper to perform the language identification and results in hallucinations occurring more often than can be tolerated. On the other hand, using faster but less accurate models is not possible as the solution must be able to run in various regions of the world, being able to differentiate between many languages. Given the issues at hand, ASR is not applicable for online usage.

What remains is selecting a state-of-the-art ASR tool to perform the speech-to-text, but only in two scenarios:
\begin{enumerate}
    \item During the training phase. As the training phase does not need to operate in real-time and is only performed once per data center, there is no need for many GPUs.
    \item Live - but only for unidentified audio files. ASR is only performed when adding new files to the database. Since adding files to the database is performed on entire early media files, not just the first few seconds of audio, the chances of wrong language identifications and hallucinations occurring are drastically reduced in this case. The cost of the system is also kept low, as most requests do not require ASR usage. Thus, several data centers can share the ASR module of the system (GPU).
\end{enumerate}

The constraints above heavily limit the options for the ASR tool, practically narrowing down the choice to Meta's MMS \cite{pratap2023scaling} or OpenAI's Whisper \cite{radford2023robust}. Of the two solutions, only Whisper provides a license suitable for proprietary usage and was thus the selected tool.

\section{Experiments}

\begin{figure*}[!b]
    \centering
    \includegraphics[width=1.0\linewidth]{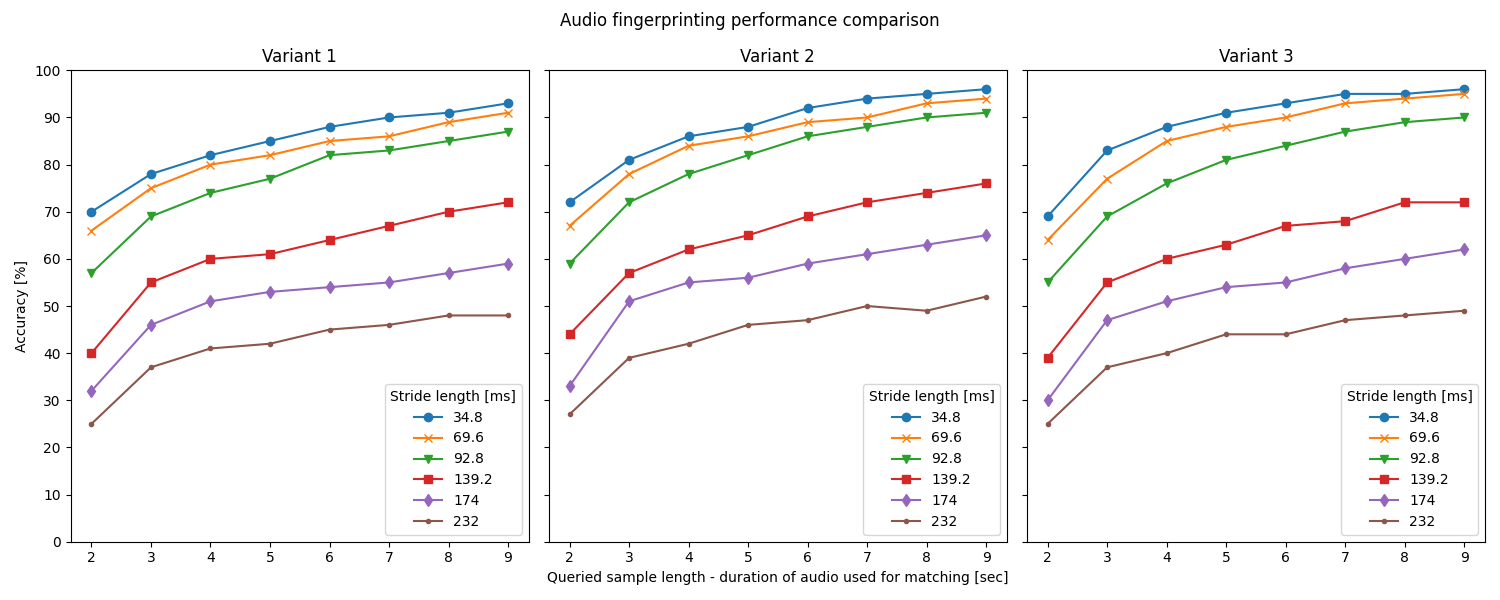}
    \caption{Accuracy of audio fingerprinting variants based on stride length and queried sample length.}
    \label{fig::accuracy}
\end{figure*}

Experiments were conducted in order to obtain better insights into the behavior of the system. Emphasis was put on the audio fingerprinting algorithm, as it is effectively the only complex part of the system which must perform in real-time.

Three variations of the audio fingerprinting component were compared:
\begin{itemize}
    \item \textit{Variant 1}: spectrogram with linearly spaced frequency bins covering the spectral range of only fundamental vocal frequencies (100Hz - 350Hz).
    \item \textit{Variant 2}: mel-scale logarithmically spaced frequency bins covering the spectral range of only fundamental vocal frequencies (100Hz - 350Hz).
    \item \textit{Variant 3}: mel-scale logarithmically spaced frequency bins covering a wider spectral range (300Hz - 2000Hz).
\end{itemize}

All three variations are heavily based on Waveprint \cite{baluja2006content,covell2007known,baluja2008waveprint}, and extract wavelets as features. For the mel-scale variants, a fixed number of 40 mel frequency bins was considered, regardless of the entire spectral range.

\subsection{Experiment Format}
An artificial dataset was used for benchmarking the problem. The dataset consists of 357 audio files, in 12 diffferent languages, with a total of 18 different speakers. The languages are very diverse: English, Japanese, Spanish, Hungarian, Arabic, etc. After creating an initial database, retrieving all of the 357 files was attempted. However, the audio samples used for retrieval were artificially deteriorated:
 \begin{itemize}
     \item Gaussian noise was added with a varying level of signal-to-noise ratio.
     \item A random offset was applied (i.e. the audio could start from any point in time). The offset sample would never exceed the length of the audio, thereby no padding was required.
     \item The playback rate was altered between 97\% and 1.03\% of normal playback rate. This is a standard modification applied to verify the robustness of the algorithm, usually by 1\%-3\% \cite{baluja2008waveprint}.
 \end{itemize}

\noindent Three main hypotheses were verified:
\begin{enumerate}
    \item After a certain threshold, increasing the length of the audio sample used for the retrieval process does not proportionally improve accuracy of retrieval.
    \item Performance is heavily impacted by the parameter known as stride length or hop length necessary for spectrogram construction. This parameter defines how slowly or quickly the spectrogram changes in time.
    \item Analyzing only a reduced vocal range of frequencies (100Hz - 350Hz) does not impact accuracy.
\end{enumerate}

\subsection{Results}
The results of the experiments are provided by Fig. \ref{fig::accuracy}. A retrieval is considered successful when the system retrieves the audio file that contains the audio sample which the system was queried with. As can be inferred from Fig. \ref{fig::accuracy}, the three hypotheses are correct.

Using 6 seconds of audio is more than enough to perform a confidential retrieval process. Increasing the length of the audio sample used for retrieval beyond 6 seconds yields diminishing returns, but impacts performance: it inherently introduces a delay required for recording more audio and it takes longer to create the fingerprints of the audio.

All variants of the fingerprinting algorithm drop in performance as the stride length increases. This is expected, as increasing the stride length during fingerprinting makes the fingerprints less robust and this is an issue most audio fingerprinting approaches share \cite{baluja2006content,covell2007known,baluja2008waveprint}.

Lastly, all three variants of the fingerprinting algorithm yield similar results. This indicates that it is not necessary to encode and fingerprint spectral content high above the fundamental frequency of human voice. Such a conclusion can yield useful for further optimizations: it can be possible to apply more "aggressive" low-pass filtering techniques, as no information will be lost if higher frequencies are altered. The results also show that using logarithmically or linearly spaced frequency bins does yield any major differences.

\section{Conclusion}

In this paper, an overview of existing audio fingerprinting techniques was reviewed, along with the current state of multilingual speech-to-text models. An interesting problem with an industrial background was considered - the problem requires real-time transcription in multiple, a priori unknown languages. After a more detailed analysis, it was concluded that the application of industry-standard automatic speech recognition tools is not applicable. Therefore, a system which relies on audio fingerprinting and fast vector search was designed. The system performs retrieval of similar speech, which was already clustered into groups based on the spoken content. This enables the system to classify audio based on speech, without the requirement of expensive hardware or recording long audio samples (necessary for better context of ASR models). Future work could investigate improving the system/approach presented, but also performing better and more detailed analysis of dedicated speech retrieval systems.

\balance

\bibliographystyle{ieeetr}
\bibliography{references.bib}

\end{document}